\documentclass[10pt,a4paper]{article}
\usepackage{jheppub_kim}
\usepackage{pdflscape}
\usepackage{amsmath}
\usepackage{amssymb}
\usepackage{dcolumn}
\usepackage{bm}
\usepackage{color}
\usepackage{epsfig}
\usepackage{amsfonts}
\usepackage{graphicx}
\usepackage{subfigure}
\usepackage{dcolumn}

\begin{document}

\title{Vaidya Spacetime in Brans-Dicke Gravity's Rainbow}



\author[a]{Prabir Rudra, }
\author[b]{~Sayani Maity}

\affiliation[a]{Department of Mathematics, Asutosh College,
Kolkata-700 026, India.}

\affiliation[b]{Department of Mathematics, Techno India Salt
Lake, Sector-V, Kolkata-700 091, India.}{}

\emailAdd{~prudra.math@gmail.com}

\emailAdd{sayani.maity88@gmail.com}

\abstract{In this note we study an energy dependent deformation of
a time dependent geometry in the background of Brans-Dicke gravity
theory. The study is performed using the gravity's rainbow
formalism. We compute the field equations in Brans-Dicke gravity's
rainbow using Vaidya metric which is a time dependent geometry. We
study a star collapsing under such conditions. Our prime objective
is to determine the nature of singularity formed as a result of
gravitational collapse and its strength.  The idea is to test the
validity of the cosmic censorship hypothesis for our model. We
have also studied the effect of such a deformation on the
thermalization process. In this regard we have calculated the
important thermodynamical quantities such as thermalization
temperature, Helmholtz free energy, specific heat and analyzed the
behavior of such quantities. }


\maketitle

\section{Introduction}

The UV completion of general relativity (GR) such that GR is
recovered in the IR limit has led to the development of
Horava-Lifshitz gravity \cite{HoravaPRD,HoravaPRL}. The concept of
different Lifshitz scaling of space and time has been used to
analyze type IIA string theory \cite{A}, type  IIB string theory
\cite{B},   AdS/CFT correspondence \cite{ho, h1, h2, oh}, dilaton
black branes \cite{d, d1}, and dilaton black holes \cite{dh, hd}.
But this is not the only way of achieving this. There is another
alternative theory where the UV completion of GR is obtained by
making the metric depend on the energy of the test particle. This
theory is termed as Gravity's Rainbow \cite{MagueijoCQG} in
literature. Although there are conceptual differences it is
believed that gravity's rainbow is related to the Horava-Lifshitz
gravity \cite{re}. This is due to the fact that both gravity's
rainbow and Horava-Lifshitz gravity are based on the modification
of the usual energy-momentum dispersion relation in the UV limit.
This modification is carried out keeping in mind that it should
reduce to the usual energy-momentum dispersion relation in the IR
limit. We know that in relativity, the form of the energy-momentum
relations are governed by the Lorentz symmetry. So it is not
strange that gravity's rainbow will disrespect such a symmetry in
the UV limit. In this connection it must be noted that in spite of
being one of the most important symmetries in nature, there are
various different quantum gravity approaches in literature which
indicates that Lorentz symmetry might only be valid at low energy
scales, and quite obviously it will breakdown in the high energy
UV limit \cite{1}-\cite{5}. Specific models where this breakdown
is expected to occur are discrete spacetime \cite{6}, string field
theory \cite{7}, spacetime foam \cite{8}, the spin-network in loop
quantum gravity (LQG)\cite{9}, non-commutative geometry \cite{10},
etc. Now such a deformation of the standard energy-momentum
dispersion relation in the UV limit of the theory will imply the
existence of a maximum energy scale. Based on the existence of
such a maximum energy scale the idea of doubly special relativity
(DSR) \cite{13} has been conceived. Gravity's rainbow is simply a
generalization of DSR applied to curved spacetime \cite{14}. As
stated earlier the geometry of the spacetime in gravity's rainbow
depend on the energy of the test particles. So it is clear that
due to such dependence each test particle of different energy will
feel a different geometry of spacetime, thus undergoing motions
differently. Thus the geometry of spacetime in gravity's rainbow
is represented by a family of energy dependent metrics forming a
rainbow of metrics. This justifies the name. In this theory the
modification in the energy-momentum dispersion relation is
introduced by energy dependent rainbow functions, $\mathcal{F}(E)$
and $\mathcal{G}(E)$, such that
\begin{equation}  \label{MDR}
E^2\mathcal{F}^2(E)-p^2\mathcal{G}^2(E)=m^2.
\end{equation}
It may be noted that here $E = E_{s}/E_P$, where $E_{s}$ is the
maximum energy that a probe in that system can take, and $E_p$ is
the Planck energy. By definition $E_s$ cannot exceed $E_p$. The
rainbow functions are chosen in such a way so that they produce
the usual energy-momentum relation of GR in the low energy IR
limit of the theory \cite{Galan:2004st}, and so they are required
to satisfy
\begin{equation}
\lim\limits_{E_s/E_P\to0} \mathcal{F}(E)=1,\qquad
\lim\limits_{E_s/E_P\to0} \mathcal{G}(E)=1.
\end{equation}
The metric in gravity's rainbow is written as
\begin{equation}  \label{rainmetric}
g^{\mu\nu}(E)=\eta^{ab}e^\mu_a(E) e^\nu_b(E).
\end{equation}

In 1961, C. H. Brans, and R. H. Dicke \cite{Brans} developed an
idea which is considered as a relativistic theory of gravitation
parallel to GR. The theory is known as the Brans-Dicke (BD) theory
of gravitation. In GR the right hand side of the field equations
consists of the stress energy tensor which is the source of the
gravitational field. But in case of BD theory the manner in which
the mass-energy pressure acts is completely different from the way
in which it acts in case of GR. In GR it is the geometric
curvature of space-time that completely controls the motion of
bodies in a gravitational field, but in case of BD theory, due to
the use of a contrasting mechanism, this dependence on geometry is
considerably reduced. These are the basic attributes that
differentiate BD theory from the traditional theories of GR. Hence
the theory demands a lot of research. Being a scalar-tensor theory
of gravitation the most important feature of BD theory is that it
consists of an additional scalar field $\phi$ which is absent in
GR. The presence of the scalar field has a strong consequence,
making the effective gravitational constant a function of space
coordinates. There is a dimensionless BD coupling constant
$\omega$ which can be tuned as per choice so as make the theory
consistent with observational evidences. This is a unique feature
of the theory and it is quite obvious that due to this provision
the theory will admit more solutions compared to GR thus enhancing
its universality. Just like GR, BD theory also predicts
gravitational deflection of light and the perihelia precession of
planets that orbit the Sun. But these phenomena totally depends on
the value of the BD parameter $\omega$ which means that it is
possible to constrain the possible values of $\omega$ from
observations of our solar system and other gravitational systems.
It is thought that GR can be obtained from the BD theory in the
limit $\omega\rightarrow \infty$ \cite{barrowmaeda}.\\

Here we will be probing the Vaidya space-time \cite{v1} in the
energy dependent deformations of BD gravity \cite{v2, v3}. In ref.
\cite{v3}, time dependent Vaidya spacetime was studied in the
background of BD gravity theory. In ref. \cite{rain1}, Rudra et al
studied the rainbow deformations of Vaidya spacetime in the
background of Galileon gravity theory and obtained interesting
results. Galileon gravity is a form of scalar tensor theory of
gravity, where there is a self-interacting term of the form
$\nabla{\phi}^{2}~ ^{\fbox{}}~\phi$, so that GR is recovered at
high densities. It contains a scalar field $\phi$ and a potential
$V(\phi)$ in its action. Since BD gravity also has a similar
set-up we are motivated to probe the energy dependent
modifications of the time dependent Vaidya space-time in its
background. This will be done via a gravitational collapse
mechanism. Nonetheless, we will also study the thermodynamical
properties of the system. It may be noted that the gravitational
collapse under different set-ups has been studied previously using
gravity's rainbow \cite{rain1}-\cite{2a}. The deformation of the
thermodynamics of black holes (BH) due to gravity's rainbow has
also been studied \cite{16,2A,4A}. The BH thermodynamics will get
modified by the rainbow functions. This is due to the fact that,
the energy $E$ which defines the rainbow functions is basically
the energy of a quantum particle near the event horizon of the BH,
emitted in the Hawking radiation. Now we can obtain a bound on
energy $E \geq 1/\Delta x $, using the uncertainty principle
$\Delta p \geq 1/\Delta x $. Furthermore, the uncertainty in
position of a particle near the event horizon can be taken to be
equal to the radius of the event horizon radius
\begin{equation}
 E\geq 1/{\Delta x} \approx 1/{r_+}.
\end{equation}
This energy bound modifies the temperature of the BH, and this
modified temperature of the BH can be used to calculate the
corrected entropy of the BH in gravity's rainbow. The energy of a
quantum particle near the event horizon is considered as the
energy of the test particle. This is the energy which is used in
defining the rainbow functions that modify the energy momentum
relations. The metric when deformed by these rainbow functions,
quite naturally deforms the BH thermodynamics. This deformation in
the thermodynamics of a BH predicts the possibility of a BH
remnant. Remnants of BH can have important implications in the
detection of mini BHs at the Large Hadron Collider (LHC)
\cite{g1}. This energy which is used in constructing rainbow
functions is dynamical in nature depending on the radial
coordinate \cite{re}. Although the explicit dependence of this
energy on the radial coordinate is unimportant for us, yet it is
important to note that the rainbow functions are dynamical in
nature, and hence cannot be gauged away by rescaling the metric.\\

Over the years gravitational collapse \cite{coll1} of stars has
been a problem of great curiosity both in classical GR as well as
modified gravity theories. The reason being that we can get at
least two types of singularities from such a phenomenon. A
singularity covered by an event horizon is a BH whereas an
uncovered singularity is popularly known as a naked singularity
(NS). Now to determine the exact initial conditions which lead to
the formation of BH or NS is a challenging astrophysical problem.
To be more precise, the quest of a physical initial condition
leading to the formation of a NS \cite{ns1,ns2,ns3} is a really
interesting problem given the validity of cosmic censorship
hypothesis (CCH) laid down by Penrose (1969) \cite{Pen} which
states that the end result of a collapsing scenario is bound to be
a singularity covered by an event horizon, i.e. a BH. Here we will
study the chosen geometry focussing ourselves on this problem.\\

The paper is organized as follows. In section 2, the field
equations for the rainbow deformed Brans-Dicke gravity are
generated. In section 3, the solution of the given system is
found. Section 4 is devoted to the study of gravitational collapse
in the system considered. In section 5, we focus ourselves on the
thermodynamical aspects of the system. Finally the paper ends with
some concluding remarks in section 6.

\section{Brans-Dicke gravity's Rainbow}

The self-interacting BD theory \cite{Brans} is described by the
following action (choosing $8\pi G=c=1$)
\begin{equation}\label{Lag}
S=\int d^{4} x \sqrt{-g}\left[\phi R- \frac{\omega(\phi)}{\phi}
{\phi}^{,\alpha} {\phi,}_{\alpha}-V(\phi)+ {\cal L}_{m}\right]
\end{equation}
where $V(\phi)$ is the self-interacting potential for the BD
scalar field $\phi$ and the constant $\omega$ is the BD parameter.

The Vaidya metric deformed by gravity's rainbow in the background
of BD theory can be given by
\begin{eqnarray*}\label{metric}
ds^{2} =-\frac{1}{\mathcal{F}^{2}(E)}\left(1-\frac{m(t,r)}{r}
\right)dt^2
+\frac{2}{\mathcal{F}(E)\mathcal{G}(E)}dtdr+\frac{1}{\mathcal{G}^{2}(E)}r^2
d\Omega_{2}^2
\end{eqnarray*}
\begin{equation}
=f(t,r)dt^2
+\frac{2}{\mathcal{F}(E)\mathcal{G}(E)}dtdr+\frac{1}{\mathcal{G}^{2}(E)}r^2
d\Omega_{2}^2,
 \end{equation}
where $\mathcal{F}(E)$ and $\mathcal{G}(E)$ are the rainbow
functions.

From the Lagrangian density given by eqn.(\ref{Lag}) we obtain the
field equations \cite{Brans}
\begin{equation}\label{field}
G_{\mu \nu}=\frac{\omega(\phi)}{{\phi}^{2}}\left[\phi  _{ , \mu}
\phi _{, \nu} - \frac{1}{2}g_{\mu \nu} \phi _{, \alpha} \phi ^{ ,
\alpha} \right] +\frac{1}{\phi}\left[\phi  _{, \mu ; \nu} -g_{\mu
\nu}~ ^{\fbox{}}~ \phi \right]-\frac{V(\phi)}{2 \phi} g_{\mu
\nu}+\frac{1}{\phi}T_{\mu \nu}
\end{equation}
and
\begin{equation}\label{tensor?!}
^{\fbox{}}~\phi=\frac{1}{3+2\omega}T-\frac{1}{3+2\omega}\left[2V(\phi)-\phi
 \frac{dV(\phi)}{d\phi}\right]
 \end{equation}
where $T=T_{\mu \nu}g^{\mu \nu}$.

Now we consider two types of fluids namely, Vaidya null radiation
and a perfect fluid having the form of the energy momentum tensor
\begin{equation}\label{collapse2.3}
T_{\mu\nu}=T_{\mu\nu}^{(n)}+T_{\mu\nu}^{(m)}
\end{equation}
with
\begin{equation}\label{collapse2.4}
T_{\mu\nu}^{(n)}=\sigma l_{\mu}l_{\nu}
\end{equation}
and
\begin{equation}\label{collapse2.5}
T_{\mu\nu}^{(m)}=(\rho+p)(l_{\mu}\eta_{\nu}+l_{\nu}\eta_{\mu})+pg_{\mu\nu}
\end{equation}
where $\rho$ and $p$ are the energy density and pressure for the
perfect fluid and $\sigma$ is the energy density corresponding to
Vaidya null radiation. In the co-moving co-ordinates
($v,r,\theta_{1},\theta_{2},...,\theta_{n}$), the two eigen
vectors of energy-momentum tensor namely $l_{\mu}$ and
$\eta_{\mu}$ are linearly independent future pointing null vectors
having components
\begin{equation}\label{collapse2.6}
l_{\mu}=(1,0,0,...,0)~~~~ and~~~~
\eta_{\mu}=\left(\frac{1}{2}\left(1-\frac{m}{r^{n-1}}\right),-1,0,...,0
\right)
\end{equation}
and they satisfy the relations
\begin{equation}\label{collapse2.7}
l_{\lambda}l^{\lambda}=\eta_{\lambda}\eta^{\lambda}=0,~
l_{\lambda}\eta^{\lambda}=-1
\end{equation}

Now, we assume the total energy-momentum tensor of the field
equation (\ref{field}) in the following form
\begin{equation}
T_{\mu\nu}=T_{\mu\nu}^{(n)}+T_{\mu\nu}^{(m)},
\end{equation}
where $T_{\mu\nu}^{(n)}$ and $T_{\mu\nu}^{(m)}$ are the
energy-momentum tensor for the Vaidya null radiation and the
energy-momentum tensor of the perfect fluid respectively and the
supporting geometry is defined as,
\begin{eqnarray}
&&T_{\mu\nu}^{(n)}=\sigma l_{\mu}l_{\nu},\nonumber\\
&&T_{\mu\nu}^{(m)}=(\rho
+p)(l_{\mu}n_{\nu}+l_{\nu}n_{\mu})+pg_{\mu\nu},
\end{eqnarray}
where $\sigma$, $\rho$ and $p$ are null radiation density, energy
density and pressure of the perfect fluid respectively. Imposing
the rainbow deformations on the linearly independent future
pointing null vectors $l_{\mu}$ and $n_{\mu}$ we get,
\begin{equation}
l_{\mu}=\left(\frac{1}{\mathcal{F}(E)},0,0,0\right)~~~~~\&~~~
n_{\mu}=\left(\frac{1}{2\mathcal{F}(E)}\left(1-\frac{m(t,r)}{r}
\right),-\frac{1}{\mathcal{G}(E)},0,0  \right)
\end{equation}
satisfying the following conditions
\begin{equation}
l_{\mu}l^{\mu}=n_{\mu}n^{\mu}=0~~~~\&~~~~l_{\mu}n^{\mu}=-1.
\end{equation}
Therefore, the non-vanishing components of the total
energy-momentum tensor will be as follows
\begin{eqnarray*}
T_{00}=\frac{\sigma}{\mathcal{F}^{2}(E)}+\frac{\rho}{\mathcal{F}^{2}(E)}\left(1-\frac{m(t,r)}{r}\right),&&
T_{01}=-\frac{\rho}{\mathcal{F}(E)\mathcal{G}(E)}, \\
\end{eqnarray*}
\begin{eqnarray}\label{energymomentum}
T_{22}=\frac{pr^2}{\mathcal{G}^2(E)}, && T_{33}=\frac{pr^2
sin^2\theta}{\mathcal{G}^2(E)}
\end{eqnarray}

The ${00}$-component of the field equations are given as
\begin{eqnarray*}
\frac{\mathcal{G}(E)\left[\mathcal{G}(E)\left(r-m\right)m'+\mathcal{F}(E)r\dot{m}\right]}{\mathcal{F}^{2}(E)r^{3}}=
\frac{\omega}{\phi^{2}}\left[\dot{\phi}^{2}+\frac{1}{2\mathcal{F}^{2}(E)}\left(1-\frac{m}{r}\right)^{2}\phi'^{2}
\mathcal{G}^{2}(E)\right]+\frac{\ddot{\phi}}{\phi}
\end{eqnarray*}

\begin{eqnarray*}
-\phi\left[\frac{\mathcal{G}(E)m}{2\mathcal{F}(E)r^{2}}-\frac{\mathcal{G}(E)m'}{2\mathcal{F}(E)r}\right]-
\phi'\left[\frac{\mathcal{G}^{2}(E)m}{2\mathcal{F}^{2}(E)r^{2}}-\frac{\mathcal{G}^{2}(E)m^{2}}
{2\mathcal{F}^{2}(E)r^{3}}-\frac{\mathcal{G}^{2}(E)m'}{2\mathcal{F}^{2}(E)r}+\frac{\mathcal{G}^{2}(E)m
m'}{2\mathcal{F}^{2}(E)r^{2}}+\frac{\mathcal{G}(E)\dot{m}}{2\mathcal{F}(E)r}\right]
\end{eqnarray*}

\begin{equation}\label{00}
+\frac{1}{\mathcal{F}^{2}(E)\phi}\left(1-\frac{m}{r}\right){\fbox{}}~\phi+\frac{1}{\mathcal{F}^{2}(E)}
\left(1-\frac{m}{r}\right)\frac{V(\phi)}{2\phi}+\frac{1}{\mathcal{F}^{2}(E)\phi}\left[\sigma+
\rho\left(1-\frac{m}{r}\right)\right]
\end{equation}

The ${11}$-component of the field equations are
\begin{equation}\label{11}
\frac{\omega}{\phi^{2}}\phi'^{2}+\frac{\phi''}{\phi}=0
\end{equation}

The ${10}$ and ${01}$-components are
\begin{eqnarray*}
-\frac{\mathcal{G}(E)m'}{\mathcal{F}(E)r^{2}}=\frac{\omega}{\phi^{2}}\left[\dot{\phi}'
-\frac{1}{2\mathcal{F}(E)\mathcal{G}(E)}\left\{\phi'^{2}\left(1-\frac{m}{r}\right)\mathcal{G}^{2}(E)\right\}\right]
+\frac{1}{\phi}\left[\dot{\phi}'+\left(\frac{\mathcal{G}(E)m}{2\mathcal{F}(E)r^{2}}
-\frac{\mathcal{G}(E)m'}{2\mathcal{F}(E)r}\right)\phi'\right.
\end{eqnarray*}
\begin{eqnarray}\label{10}
\left.-\frac{1}{\mathcal{F}(E)\mathcal{G}(E)}{\fbox{}}~\phi\right]-\frac{V(\phi)}{2\phi\mathcal{F}(E)\mathcal{G}(E)}
-\frac{\rho}{\phi\mathcal{F}(E)\mathcal{G}(E)}
\end{eqnarray}

Finally the ${22}$ and ${33}$-components are given by
\begin{equation}\label{2233}
\frac{1}{2}rm''=\frac{\omega
r^{2}}{2\phi^{2}}\left(1-\frac{m}{r}\right)\phi'^{2}-\frac{1}{\phi}\left[\frac{\mathcal{F}(E)r}{\mathcal{G}(E)}
\dot{\phi}+\left(r-m\right)\phi'-\frac{r^{2}}{\mathcal{G}^{2}(E)}{\fbox{}}~\phi\right]+\frac{V(\phi)r^{2}}
{2\phi\mathcal{G}^{2}(E)}-\frac{pr^{2}}{\mathcal{G}^{2}(E)\phi}
\end{equation}
where ${\fbox{}}~\phi$ is given by,
\begin{equation}\label{box}
{\fbox{}}~\phi=\mathcal{F}(E)\mathcal{G}(E)\dot{\phi}'+\frac{\mathcal{G}^{2}(E)}{2r}\left(1-m'\right)\phi'+
\frac{\mathcal{F}(E)\mathcal{G}(E)}{2r}\dot{\phi}+\left(1-\frac{m}{r}\right)\mathcal{G}^{2}(E)\phi''
\end{equation}
Here dot and dash represents derivative with respect to $t$ and
$r$ respectively.

\section{The Solution}
In this section we will find the solutions of the field equations
given in the previous section. From equation(\ref{11}) we get,
\begin{equation}\label{phi}
\phi(r,t)=\left[r+r\omega-f_{1}(t)\right]^{\frac{1}{1+\omega}}f_{2}(t)
\end{equation}
where $f_{1}(t)$ and $f_{2}(t)$ are arbitrary functions of time.
We assume that the matter field follows the barotropic equation of
state given by,
\begin{equation}\label{pressure}
p=k\rho
\end{equation}
Using the equations (\ref{10}), (\ref{2233}), (\ref{box}),
(\ref{phi}) and (\ref{pressure}) we get the following differential
equation for the graviton mass $m$,
\begin{eqnarray*}\label{mass}
\left[\frac{\mathcal{G}(E)}{2kr\mathcal{F}(E)}\right]m''+\left[\frac{\mathcal{G}(E)}{2k\mathcal{F}(E)r}\frac{\phi'}
{\phi}+\frac{\mathcal{G}(E)}{\mathcal{F}(E)r^{2}}\right]m'+\left[\frac{\left(k+1\right)\omega
\mathcal{G}(E)}{2k\mathcal{F}(E)r}\left(\frac{\phi'}{\phi}\right)^{2}+\frac{\mathcal{G}(E)\left(k-2\right)}
{2k\mathcal{F}(E)r^{2}}\frac{\phi'}{\phi}\right.
\end{eqnarray*}

\begin{eqnarray*}
\left.+\frac{\mathcal{G}(E)\left(k+1\right)}{kr\mathcal{F}(E)}\frac{\phi''}{\phi}\right]m+
\left[\left(\frac{\omega}{\phi}-\frac{1}{k}\right)\frac{\dot{\phi}'}{\phi}-
\frac{\mathcal{G}(E)}{\mathcal{F}(E)}\frac{\left(k+1\right)}{k}\frac{\phi''}{\phi}-
\frac{\omega\left(k+1\right)\mathcal{G}(E)}{2\mathcal{F}(E)k}\left(\frac{\phi'}{\phi}\right)^{2}\right.
\end{eqnarray*}
\begin{eqnarray}
\left.+\frac{\mathcal{G}(E)\left(1-k\right)}{2k\mathcal{F}(E)r}\frac{\phi'}{\phi}+\left(\frac{1-k}{2kr}\right)
\frac{\dot{\phi}}{\phi}+\frac{2-k\phi}{2k\mathcal{F}(E)\mathcal{G}(E)}\frac{V(\phi)}{\phi^{2}}\right]=0
\end{eqnarray}
Unfortunately due to high complexity, a general solution for the
above differential equation cannot be obtained by the known
mathematical methods. So we seek solutions for special cases. We
see that if we consider $\phi(r,t)$ in a form where the variable
$r$ and $t$ can be separated, we can put the equation in the
Cauchy-Euler form from where we can get a solution. So to
facilitate further computations we consider $f_{1}(t)=0$ in
equation (\ref{phi}). So the expression for $\phi$ takes the form
\begin{equation}\label{separated}
\phi(r,t)=\left(r+r\omega\right)^{\frac{1}{1+\omega}}f_{2}(t)
\end{equation}
Obviously it must be admitted that this assumption produces a
particular class of solution of the collapsing system and not the
general solution. But this class of solution is of interest to us
as far as the mathematical integrity of the problem is concerned.
Now using eqn.(\ref{separated}) in eqn.(\ref{mass}) we get the
following differential equation,
\begin{eqnarray*}
r^{2}m''+\left[\frac{1}{1+\omega}+2k\right]rm'+\left[\frac{k-3\omega-2}{\left(1+\omega\right)^{2}}\right]m=
\left[\frac{k-2\omega-1}{\left(1+\omega\right)^{2}}\right]r-\frac{2k\mathcal{F}(E)}{\mathcal{G}(E)}\frac{\dot{f_{2}(t)}}{f_{2}(t)}\times
\end{eqnarray*}
\begin{eqnarray}
\left[\frac{1}{1+\omega}\left\{\frac{\omega}
{\left(1+\omega\right)^{\frac{1}{1+\omega}}f_{2}(t)}-\frac{1}{k}\right\}+\frac{1-k}{2k}\right]
r^{2}-\left[\frac{2-k\left(1+\omega\right)^{\frac{1}{1+\omega}}f_{2}(t)V(\phi)}
{\mathcal{G}^{2}(E)\left(1+\omega\right)^{\frac{2}{1+\omega}}\left(f_{2}(t)\right)^{2}}\right]r^{3}
\end{eqnarray}
We solve the above equation and get the following solution for
$m$,
\begin{eqnarray*}\label{massfinal}
m(t,r)=f_{3}(t)r^{\omega_{1}}+f_{4}(t)r^{\omega_{2}}+\frac{\left(k-2\omega-1\right)r}{\left(1+\omega\right)^{2}
\left(1-\omega_{1}\right)\left(1-\omega_{2}\right)}-\frac{2k\mathcal{F}(E)}{\mathcal{G}(E)}\frac{\dot{f_{2}(t)}}
{f_{2}(t)}\frac{r^{2}}{\left(2-\omega_{1}\right)\left(2-\omega_{2}\right)}\times
\end{eqnarray*}
\begin{eqnarray}\label{m}
\left\{\frac{1}{1+\omega}\left(\frac{\omega}{\left(1+\omega\right)^{\frac{1}{1+\omega}}f_{2}(t)}
-\frac{1}{k}\right)+\frac{1-k}{2k}\right\}-\frac{2-k\left(1+\omega\right)^{\frac{1}{1+\omega}}f_{2}(t)V(\phi)}{\mathcal{G}(E)^{2}\left(1+\omega\right)
^{\frac{2}{1+\omega}}\left\{f_{2}(t)\right\}^{2}}\frac{r^{3}}{\left(3-\omega_{1}\right)\left(3-\omega_{2}\right)}
\end{eqnarray}
where $\omega_{1}$ and $\omega_{2}$ are given by,
\begin{equation}
\omega_{1},\omega_{2}=\frac{1}{2\left(1+\omega\right)}\left[\left\{-2k+\left(1-2k\right)\omega\right\}\pm
\sqrt{\left\{2k+\omega\left(2k-1\right)\right\}^{2}-4\left(k-3\omega-2\right)}\right]
\end{equation}
From the above relation it is seen that the admissible range of
the barotropic parameter $k$ is given by
\begin{equation}
k\in \left(-\infty~,
\frac{1+\omega+\omega^{2}-\sqrt{-7-26\omega-30\omega^{2}-12\omega^{3}}}{2\left(1+2\omega+\omega^{2}\right)}\right]
\bigcup
\left[\frac{1+\omega+\omega^{2}+\sqrt{-7-26\omega-30\omega^{2}-12\omega^{3}}}{2\left(1+2\omega+\omega^{2}\right)},
\infty\right)
\end{equation}
Using equation(\ref{massfinal}) in equation(\ref{metric}) we get
the rainbow deformed Vaidya metric in BD gravity as follows,
\begin{eqnarray*}
ds^{2}
=-\frac{1}{\mathcal{F}^{2}(E)}\left(1-f_{3}(t)r^{\omega_{1}-1}-f_{4}(t)r^{\omega_{2}-1}-\frac{\left(k-2\omega-1\right)}
{\left(1+\omega\right)^{2}
\left(1-\omega_{1}\right)\left(1-\omega_{2}\right)}+\frac{2k\mathcal{F}(E)}{\mathcal{G}(E)}\frac{\dot{f_{2}(t)}}{f_{2}(t)}\frac{r}{\left(2-\omega_{1}\right)
\left(2-\omega_{2}\right)}\right.
\end{eqnarray*}

\begin{eqnarray*}
\left.\times\left\{\frac{1}{1+\omega}\left(\frac{\omega}{\left(1+\omega\right)^{\frac{1}{1+\omega}}f_{2}(t)}
-\frac{1}{k}\right)+\frac{1-k}{2k}\right\}+\left\{\frac{2-k\left(1+\omega\right)^{\frac{1}{1+\omega}}f_{2}(t)V(\phi)}{\mathcal{G}(E)^{2}
\left(1+\omega\right)
^{\frac{2}{1+\omega}}\left\{f_{2}(t)\right\}^{2}}\right\}\frac{r^{2}}{\left(3-\omega_{1}\right)\left(3-\omega_{2}\right)}
\right)dt^2
\end{eqnarray*}
\begin{eqnarray}
+\frac{2}{\mathcal{F}(E)\mathcal{G}(E)}dtdr+\frac{1}{\mathcal{G}^{2}(E)}r^2
d\Omega_{2}^2
\end{eqnarray}

\section{Gravitational Collapse}

In this section, we use the concept of radial null geodesics to
explore the existence of NS in generalized Vaidya space-time. We
need first to check whether it is possible to have outgoing radial
null geodesics that were terminated in the past at the central
singularity $r=0$. The type of the singularity (NS or BH) can be
determined by the existence of radial null geodesics emerging from
the singularity. The singularity is said to be locally naked if
there exist such geodesics and is said to be BH if geodesics do
not exist. The catastrophic gravitational collapse causes two
possible types of singularities which could be NS or a BH.
Although CCH states that, a gravitational collapse always results
in a BH, yet there is no rigorous proof for that. We have already
seen that inhomogeneous dust cloud may result in a NS through a
collapse \cite{Eardley1}. Fluids with different equations of state
other than dust also give rise to considerable results
\cite{Joshi1}. So the validity of the hypothesis is quite
questionable. At least keeping the above literature in view the
censorship hypothesis needs to get generalized \cite{Joshi2}.

We assume that $R(t, ~r)$ is the physical radius at time $t$ of
the shell labelled by $r$. At the starting epoch $t=0$ we should
have $R(0,~r)=r$. In the inhomogeneous case, different shells
could become singular at different times. Now if there are future
directed radial null geodesics emanating out of the singularity,
with a well defined tangent at  the singularity $\frac{dR}{dr}$
must tend to a finite limit in the limit of approach to the
singularity in the past along these trajectories. When reaching
the points $(t_0, ~r)=(t_{0}, ~0)$, the singularity $R(t_0,0)=0$
occurs which corresponds to the physical situation where matter
shells are crushed to zero radius. This type of singularity
($r=0$) is called a central singularity. The singularity is a NS
if there exists future directed non-space like curves in the space
time with their past end points rooted in the singularity. Now if
the outgoing null geodesics are traced back so as they terminate
in the past at the central singularity ($r=0$ at $t=t_0$) where
$R(t_0, 0)=0$, then along these geodesics we should have
$R\rightarrow 0$ as $r\rightarrow 0$ \cite{Singh1}.

\begin{figure}
~~~~~\includegraphics[height=2.5in,width=2.5in]{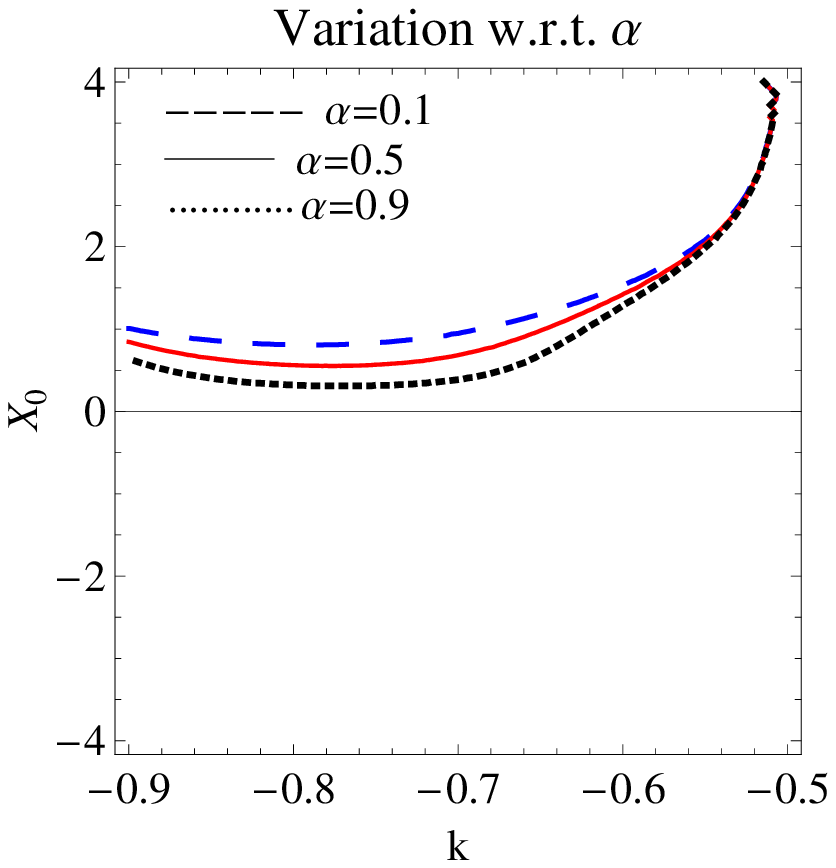}~~~~~~~~~~~~~\includegraphics[height=2.5in,width=2.5in]{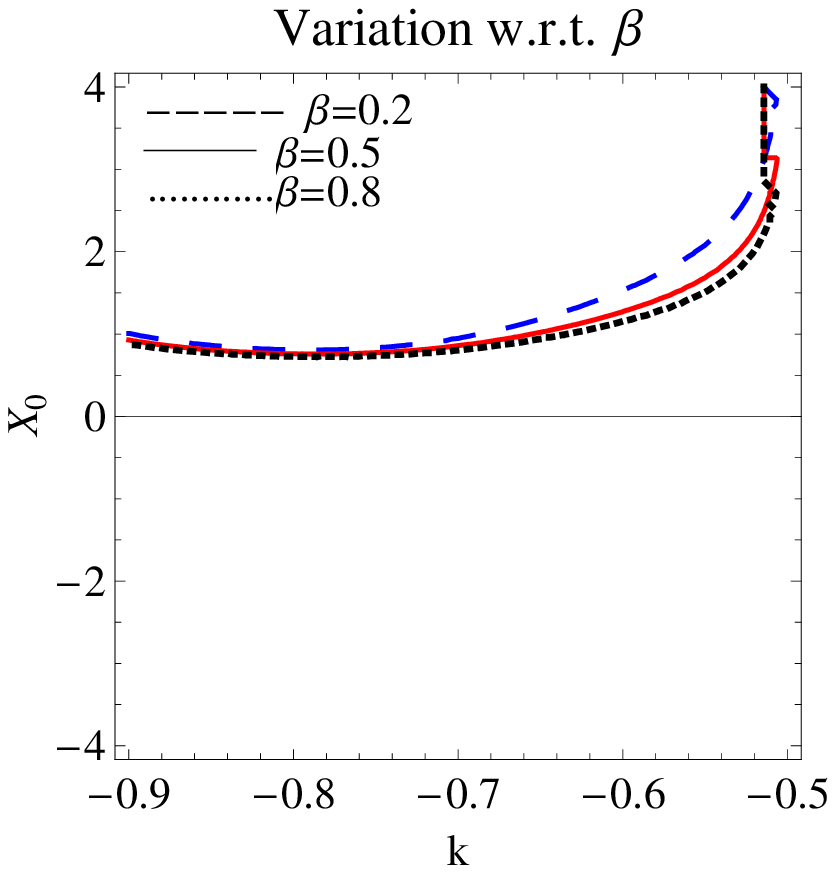}~~~~~~~\\\\

~~~~~~~~~~~~~~~~~~~~~~~~~~~~~~~~Fig.1~~~~~~~~~~~~~~~~~~~~~~~~~~~~~~~~~~~~~~~~~~~~~~~~~~~~~~~Fig.2~~~~~~~~~\\

\vspace{2mm} \textit{\textbf{Figs 1 and 2} show the variation of
$X_{0}$ with $k$ for different values of $\alpha$ and $\beta$
respectively in Brans-Dicke gravity's rainbow.\\\\ In fig.1 the
other parameters are fixed at $\beta=0.2$, $\gamma=5$,
$V_{0}=0.1$, $\omega=-0.5$, $\eta=1$, $E_{1}=1.42 \times
10^{-13}$, $E_{p}=1.221 \times 10^{19}$.\\\\ In fig.2 the other
parameters are taken as $\alpha=0.1$, $\gamma=5$, $V_{0}=0.1$,
$\omega=-0.5$, $\eta=1$, $E_{1}=1.42 \times 10^{-13}$,
$E_{p}=1.221 \times 10^{19}$.}
\end{figure}
\begin{figure}
~~~~~~~~~~\includegraphics[height=2.5in, width=2.5in]{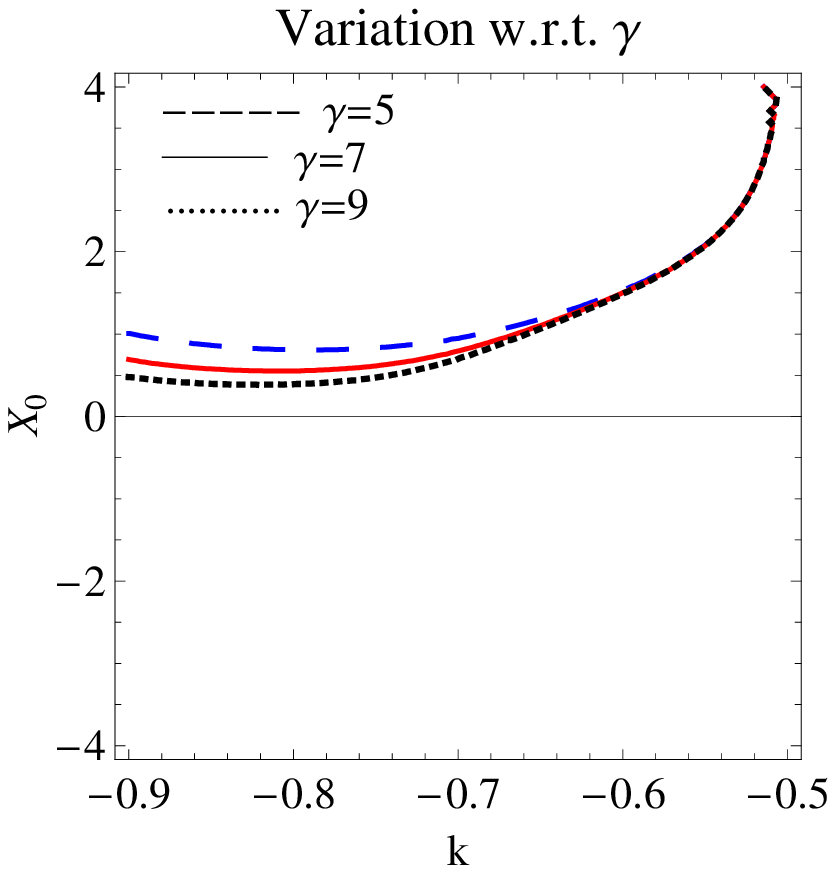}~~~~~~~~~~~~~\includegraphics[height=2.5in,width=2.5in]{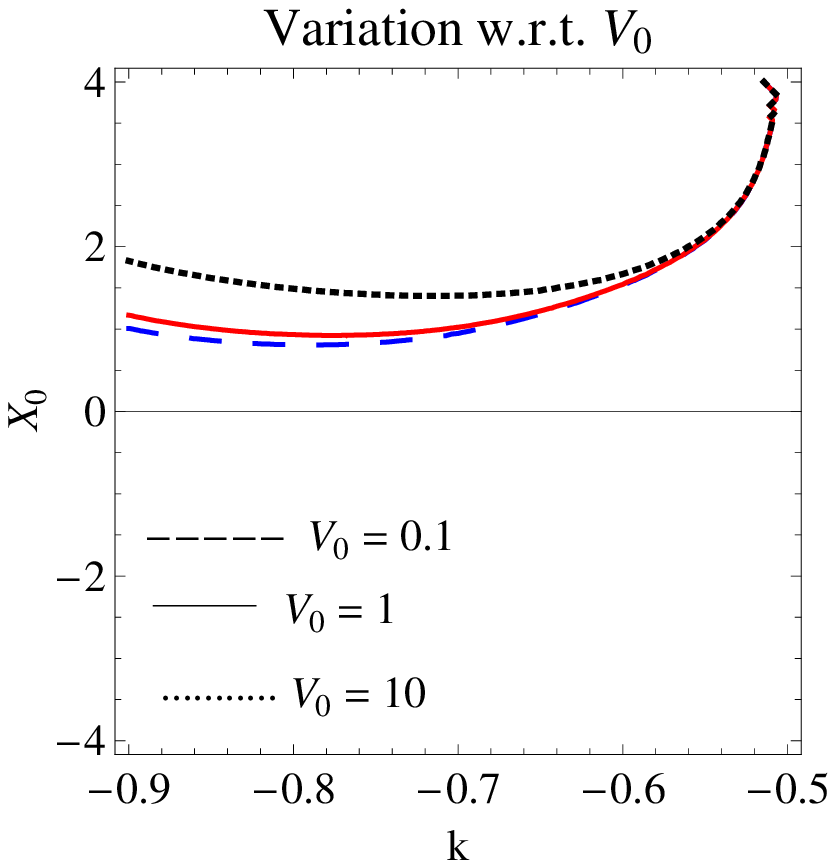}~~~~~~~\\\\

~~~~~~~~~~~~~~~~~~~~~~~~~~~~~~~Fig.3~~~~~~~~~~~~~~~~~~~~~~~~~~~~~~~~~~~~~~~~~~~~~~~~~~~~~~~~~~~~~~~~Fig.4~~~~~~~~\\

\vspace{2mm} \textit{\textbf{Figs 3 and 4} show the variation of
$X_{0}$ with $k$ for different values of $\gamma$ and $V_{0}$
respectively in Brans-Dicke gravity's rainbow.\\\\ In fig.3 the
other parameters are fixed at $\alpha=0.1, \beta=0.2$,
$V_{0}=0.1$, $\omega=-0.5$, $\eta=1$, $E_{1}=1.42 \times
10^{-13}$, $E_{p}=1.221 \times 10^{19}$.\\\\ In fig.4 the other
parameters are taken as $\alpha=0.1, \beta=0.2$, $\gamma=5$,
$\omega=-0.5$, $\eta=1$, $E_{1}=1.42 \times 10^{-13}$,
$E_{p}=1.221 \times 10^{19}$.}
\end{figure}

\begin{figure}
~~~~~~~~~~~~~~~~~~~~~~~\includegraphics[height=2.5in, width=2.5in]{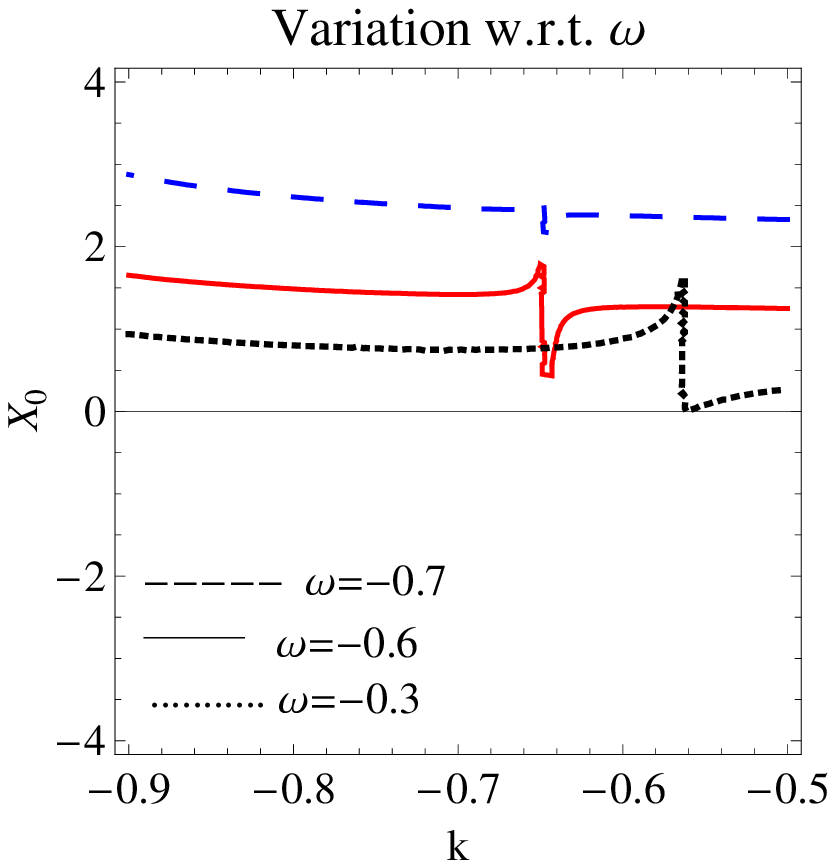}~~~~~~~\\\\

~~~~~~~~~~~~~~~~~~~~~~~~~~~~~~~~~~~~~~~~~~~~~~Fig.5~~~~~~~~~~~~~~~~~~~~~~~~~~~~~~\\

\vspace{2mm} \textit{\textbf{Fig 5} show the variation of $X_{0}$
with $k$ for different values of $\omega$ in Brans-Dicke gravity's rainbow.\\\\
In fig.5 the other parameters are fixed at $\alpha=0.1, \beta=5$,
$\gamma=1$, $V_{0}=0.1$, $\eta=1$, $E_{1}=1.42 \times 10^{-13}$,
$E_{p}=1.221 \times 10^{19}$.}

\end{figure}
The equation for outgoing radial null geodesics can be obtained
from equation (\ref{metric}) by putting $ds^{2}=0$ and
$d\Omega_{2}^{2}=0$ as
\begin{equation}
\frac{dt}{dr}=\frac{2\mathcal{F}(E)}{\mathcal{G}(E)\left(1-\frac{m(t,r)}{r}\right)}.
\end{equation}
From the above expression it is quite clear that at $r=0,~t=0$
there is a singularity of the above differential equation. Suppose
we consider a parameter $X=\frac{t}{r}$. Using this parameter we
can study the limiting behavior of the function $X$ as we approach
the singularity at $r=0,~t=0$ along the radial null geodesic. If
we denote the limiting value by $X_{0}$ then
\begin{eqnarray}\label{X0}
\begin{array}{c}
X_{0}\\\\
{}
\end{array}
\begin{array}{c}
=lim~~ X \\
\begin{tiny}t\rightarrow 0\end{tiny}\\
\begin{tiny}r\rightarrow 0\end{tiny}
\end{array}
\begin{array}{c}
=lim~~ \frac{t}{r} \\
\begin{tiny}t\rightarrow 0\end{tiny}\\
\begin{tiny}r\rightarrow 0\end{tiny}
\end{array}
\begin{array}{c}
=lim~~ \frac{dt}{dr} \\
\begin{tiny}t\rightarrow 0\end{tiny}\\
\begin{tiny}r\rightarrow 0\end{tiny}
\end{array}
\begin{array}{c}
=lim~~ \frac{2\mathcal{F}(E)}{\mathcal{G}(E)\left(1-\frac{m(t,r)}{r}\right)} \\
\begin{tiny}t\rightarrow 0\end{tiny}~~~~~~~~~~~~\\
\begin{tiny}r\rightarrow 0\end{tiny}~~~~~~~~~~~~
 {}
\end{array}
\end{eqnarray}
Using equations (\ref{m}) and (\ref{X0}), we have
\begin{eqnarray*}
\frac{2}{X_{0}}=
\begin{array}llim\\
\begin{tiny}t\rightarrow 0\end{tiny}\\
\begin{tiny}r\rightarrow 0\end{tiny}
\end{array}
\frac{\mathcal{G}(E)}{\mathcal{F}(E)}\left[1-f_{3}(t)r^{\omega_{1}-1}-f_{4}(t)r^{\omega_{2}-1}-\frac{\left(k-2\omega-1\right)}
{\left(1+\omega\right)^{2}\left(1-\omega_{1}\right)\left(1-\omega_{2}\right)}+\frac{2k\mathcal{F}(E)}
{\mathcal{G}(E)}\frac{r}{\left(2-\omega_{1}\right)\left(2-\omega_{2}\right)}\right.
\end{eqnarray*}
\begin{eqnarray}
\left.\times\frac{\dot{f_{2}(t)}}{f_{2}(t)}\left\{\frac{1}{1+\omega}\left(\frac{\omega}{\left(1+\omega\right)^{\frac{1}{1+\omega}}f_{2}(t)}
-\frac{1}{k}\right)+\frac{1-k}{2k}\right\}+\left\{\frac{2-k\left(1+\omega\right)^{\frac{1}{1+\omega}}f_{2}(t)V(\phi)}{\mathcal{G}(E)^{2}
\left(1+\omega\right)
^{\frac{2}{1+\omega}}\left\{f_{2}(t)\right\}^{2}}\right\}\frac{r^{2}}{\left(3-\omega_{1}\right)\left(3-\omega_{2}\right)}\right]
\end{eqnarray}
Here we will take the potential $V(\phi)$ in the power law form,
i.e., $V(\phi)=V_{0}\phi^{n}$, $n$ and $V_{0}$ being a real
number. Now choosing $f_{2}(t)=\gamma t$,~~$f_{3}(t)=\alpha
t^{1-\omega_{1}}$,~~$f_{4}(t)=\beta t^{1-\omega_{2}}$~~and ~$n=-1$
we obtain an algebraic equation for $X_{0}$ as
\begin{eqnarray*}
\alpha X_{0}^{3-\omega_{1}}+\beta
X_{0}^{3-\omega_{2}}+\frac{\left(k-2\omega-1\right)}{\left(1+\omega\right)^{2}\left(1-\omega_{1}\right)
\left(1-\omega_{2}\right)}X_{0}^{2}+\left[\frac{\mathcal{F}(E)}{\mathcal{G}(E)}
\left\{2-\frac{\omega-k\left(1+\omega\right)-1}{\mathcal{G}(E)\left(2-\omega_{1}\right)
\left(2-\omega_{2}\right)\left(1+\omega\right)}\right\}-1\right]X_{0}
\end{eqnarray*}
\begin{eqnarray}
-\frac{\left(2-kV_{0}\right)}{\mathcal{G}^{2}(E)\left(1+\omega\right)^{\frac{2}{1+\omega}}\gamma^{2}
\left(3-\omega_{1}\right)\left(3-\omega_{2}\right)}=0
\end{eqnarray}
where $\alpha$, $\beta$ and $\gamma$ are arbitrary constants. It
must be mentioned over here that the choices of $f_{2}(t)$,
$f_{3}(t)$ and $f_{4}(t)$ are somewhat self-similar in nature. The
choices have been made depending on the definition of $X_{0}$ in
equation (\ref{X0}) such that the ratio $t/r$ can be formed. Non
self similar assumptions can also be made, but that will result in
either removal of terms or creation of mathematically undefined
terms. As a result of this a lot of information about the system
will be lost which is undesirable.

Now if we get only non-positive solution of the equation we can
assure the formation of a BH. Getting a positive root indicates a
chance to get a NS. Since the obtained equation is a highly
complicated one, it is extremely difficult to find out an analytic
solution of $X_{0}$ in terms of the variables involved. So our
idea is to find out different numerical solutions of $X_{0}$, by
assigning particular numerical values to the associated
parameters, i.e., $\alpha$, $\beta$, $\gamma$, $V_{0}$, $k$ and
$\omega$.

\subsubsection{Numerical Analysis}

Since there are many parameters to deal with, we have generated
plots for the function $X_{0}$ by varying a particular parameter
and fixing others. This helps in understanding the dependencies
effectively. Since the evolution of universe and its different
phases are characterized by the value of the equation of state
$k$, in figs.1 to 5, we have obtained the profiles for the
variable $X_{0}$ with respect to the barotropic EoS parameter $k$.
Motivated from refs. \cite{Amelino1, Amelino2}, we have used the
following rainbow functions,
\begin{equation}
\mathcal{F}(E)=1,~~~~~\mathcal{G}(E)=\sqrt{1-\eta\left(\frac{E_{1}}{E_{p}}\right)}
\end{equation}
In the above expressions, $E_{p}$ is the planck energy given by
$E_{p}=1/\sqrt{G}=1.221 \times 10^{19}$ GeV, where $G$ is the
gravitational constant and $E_{1}=1.42 \times 10^{-13}$
\cite{Amelino1, Amelino2}. In ref. \cite{Amelino2}, the value of
$\eta$ has been roughly computed as $\eta\approx 1$, following
which we have used $\eta=1$ in our study .

From all the five plots we see that the trajectories for $X_{0}$
appear in the positive level, thus ruling out the possibility of
formation of BH as an end state of collapse. This is a
counter-example of Cosmic censorship hypothesis. In fig.1,
$k-X_{0}$ plots have been obtained for different values of
parameter $\alpha$. We see that with the increase in the value of
$\alpha$, the trajectories push towards the $k$-axis, thus
exhibiting a reduced tendency of formation of NS. We get a similar
scenario when $\beta$ and $\gamma$ is varied in figs.2 and 3
respectively. In fig.4, $k-X_{0}$ trajectories are obtained for
different values of the field potential parameter $V_{0}$. Here we
see a reversed result. With the increase in the value of $V_{0}$
the $X_{0}$ profiles tend towards higher positive range, thus
decreasing the tendency of BH formation. Finally in fig.5, we
obtained plots for variable values of BD parameter $\omega$. Here
an increase in the value of the $\omega$ parameter decreases the
tendency of NS formation mimicking the first three cases. We know
that in the limit $\omega \rightarrow \infty$, GR is recovered
from the BD gravity. So here we can see that in the limit when the
theory tends towards GR, the tendency of formation of BH increase.
This shows that there is a greater tendency of the cosmic
censorship hypothesis to be true in case of GR. But as the gravity
is modified, with greater deviations the hypothesis loses its
significance and we get counter-examples as in the present work.
As this is our prime motivation, we have worked with small values
of $\omega$ so that we can study the scenarios with greater
deviations from GR.

\subsubsection{Strength of Singularity}

The strength of singularity is defined as the measure of its
destructive capacity. The prime concern is that whether extension
of space-time is possible through the singularity or not under any
situation. Following Tipler \cite{Tipler} a curvature singularity
is said to be strong if any object hitting it is crushed to zero
volume. In \cite{Tipler} the condition for a strong singularity is
given by,
\begin{eqnarray}
\begin{array}{c}
S=lim~~ \tau^{2}\psi \\
\begin{tiny}\tau\rightarrow 0\end{tiny}\\
\end{array}
\begin{array}{c}
=lim~~ \tau^{2}R_{\mu\nu}K^{\mu}K^{\nu}>0 \\
\begin{tiny}\tau\rightarrow 0\end{tiny}\\
\end{array}
\end{eqnarray}
where $R_{\mu\nu}$ is the Ricci tensor, $\psi$ is a scalar given
by $\psi=R_{\mu\nu}K^{\mu}K^{\nu}$, where
$K^{\mu}=\frac{dx^{\mu}}{d\tau}$ is the tangent to the non
spacelike geodesics at the singularity and $\tau$ is the affine
parameter. In the paper \cite{Maharaj} Mkenyeleye et al have shown
that,
\begin{eqnarray}\label{maha}
\begin{array}{c}
S=lim~~ \tau^{2}\psi \\
\begin{tiny}\tau\rightarrow 0\end{tiny}\\
\end{array}
\begin{array}{c}\label{stren}
=\frac{1}{4}X_{0}^{2}\left(2\dot{m_{0}}\right) \\
\begin{tiny}~\end{tiny}\\
\end{array}
\end{eqnarray}
where
\begin{eqnarray}
\begin{array}{c}
m_{0}=lim~~ m(t,r) \\
\begin{tiny}t\rightarrow 0\end{tiny}\\
\begin{tiny}r\rightarrow 0\end{tiny}
\end{array}
\end{eqnarray}
and
\begin{eqnarray}\label{massd}
\begin{array}{c}
\dot{m_{0}}=lim~~ \frac{\partial}{\partial~t}\left(m(t,r)\right) \\
\begin{tiny}t\rightarrow 0\end{tiny}\\
\begin{tiny}r\rightarrow 0\end{tiny}
\end{array}
\end{eqnarray}

Using eqn.(\ref{massfinal}) in the above relation (\ref{maha}) we
get
\begin{eqnarray}\label{strength2}
\begin{array}{c}
S=lim~~ \tau^{2}\psi \\
\begin{tiny}\tau\rightarrow 0\end{tiny}\\
\end{array}
\begin{array}{c}
=\frac{1}{2}X_{0}^{2}\left[\alpha\left(1-\omega_{1}\right)X_{0}^{-\omega_{1}}+\beta\left(1-\omega_{2}\right)X_{0}^{-\omega_{2}}
+\frac{2k\mathcal{F}(E)}{\mathcal{G}(E)\left(2-\omega_{1}\right)\left(2-\omega_{2}\right)}\left\{\frac{2\omega}{\gamma
\left(1+\omega\right)^{\frac{2+\omega}{1+\omega}}}+\frac{1-k}{2k}-\frac{1}{k\left(1+\omega\right)}\right\}\frac{1}{X_{0}^{2}}\right] \\
\begin{tiny}~\end{tiny}\\
\end{array}
\end{eqnarray}
In the paper \cite{Maharaj} it has also been shown that the
relation between $X_{0}$ and the limiting values of mass is given
by,
\begin{equation}\label{xmass}
X_{0}=\frac{2}{1-2m_{0}'-2\dot{m_{0}}X_{0}}
\end{equation}
where
\begin{eqnarray}\label{dashedmass}
\begin{array}{c}
m_{0}'=lim~~ \frac{\partial}{\partial~r}\left(m(t,r)\right) \\
\begin{tiny}t\rightarrow 0\end{tiny}\\
\begin{tiny}r\rightarrow 0\end{tiny}
\end{array}
\end{eqnarray}
and $\dot{m_{0}}$ is given by the eqn.(\ref{massd}). Using
eqns.(\ref{massfinal}), (\ref{massd}) and (\ref{dashedmass}) in
eqn.(\ref{xmass}) we get an equation for $X_{0}$ which can be
solved to check the existence of positive roots. The existence of
such a positive root signifies that the singularity is naked.
Using these positive values of $X_{0}$ in the
eqn.(\ref{strength2}) we get the conditions for which $S=lim~~
\tau^{2}\psi >0$, which gives the conditions under which we get a
strong naked singularity.


\section{Thermodynamics}

In this section, we would like to focus on the thermodynamical
aspects of Vaidya spacetime in BD gravity's rainbow. To
investigate the effect of such a spacetime on the thermalization
process, here we consider the the thermalization temperature by
the following relation \cite{1ECAKDLY14},
\begin{equation}\label{temp}
T=\frac{1}{4\pi}\frac{d}{dr}f(t,r)|_{r=r_{h}},
\end{equation}
where $r_{h}$ is the event horizon obtained from the relation $f(t,r)=0$, i.e,
\begin{eqnarray*}
-\frac{1}{\mathcal{F}^{2}(E)}\left(1-f_{3}(t)r^{\omega_{1}-1}-f_{4}(t)r^{\omega_{2}-1}-\frac{\left(k-2\omega-1\right)}
{\left(1+\omega\right)^{2}
\left(1-\omega_{1}\right)\left(1-\omega_{2}\right)}+\frac{2k\mathcal{F}(E)}{\mathcal{G}(E)}\frac{\dot{f_{2}(t)}}{f_{2}(t)}\frac{r}{\left(2-\omega_{1}\right)
\left(2-\omega_{2}\right)}\right.
\end{eqnarray*}

\begin{equation}
\left.\times\left\{\frac{1}{1+\omega}\left(\frac{\omega}{\left(1+\omega\right)^{\frac{1}{1+\omega}}f_{2}(t)}
-\frac{1}{k}\right)+\frac{1-k}{2k}\right\}+\left\{\frac{2-k\left(1+\omega\right)^{\frac{1}{1+\omega}}f_{2}(t)V(\phi)}{\mathcal{G}(E)^{2}
\left(1+\omega\right)
^{\frac{2}{1+\omega}}\left\{f_{2}(t)\right\}^{2}}\right\}\frac{r^{2}}{\left(3-\omega_{1}\right)\left(3-\omega_{2}\right)}
\right)=0.
\end{equation}
The real positive root of the above equation describes the radius
of the event horizon. Figure 6 presents the typical behavior of
$f(t,r)$ in terms of $r$ for different values of the parameter
$k$. Here we have used the values of the other parameters such as
$E_{1}, E_{P}, n, F, V_{0}, f_{2}(t), f_{3}(t), f_{4}(t)$ as
described in section 4. Figure 6(a) shows the horizon structure of
the Vaidya spacetime in BD gravity's rainbow and in 6(b) we have
shown the zoomed range of outer horizon obtained from plot
6(a).These two figures yield $r_{h}\approx1$ for the selected
value of the parameters. In \cite{1YHPRBPMFAFFD17} a time
dependent geometry in massive theory of gravity has been analyzed
and thermodynamical aspect of such geometry has been studied.

Thermalization temperature given by equation (\ref{temp}), due to
the Vaidya spacetime in BD gravity's rainbow, takes the form
\begin{eqnarray*}\label{temperature}
T=\frac{1}{4\pi
\mathcal{F}(E)}\Big{[}r^{-2+\omega_{1}}f_{3}(t)(\omega_{1}-1)+\frac{1}{\mathcal{G}(E)f_{2}(t)^{2}}\Big{(}-2(1+\omega)^{\frac{-2}{(1+\omega)}}\left(
\frac{2r}{(\omega_{1}-3)(\omega_{2}-3)}+\frac{k\dot{f_{2}(t)}\omega(1+\omega)^{-1+\frac{1}{1+\omega}}}{(\omega_{1}-2)(\omega_{2}-2)}\right)
\end{eqnarray*}
\begin{equation}
+f_{2}(t)\left(\frac{k(2+n+2\omega)V_{0}r^{n+1}f_{2}(t)^{n}(1+\omega)^{\frac{n-2-\omega}{1+\omega}}}{(\omega_{1}-3)(\omega_{2}-3)}
+\frac{\dot{f_{2}(t)}(1+k+(k-1)\omega)}{(1+\omega)(\omega_{1}-2)(\omega_{2}-2)}\right)+f_{4}(t)(\omega_{2}-1)r^{-2+\omega_{2}}\Big{)}\Big{]}.
\end{equation}

\begin{figure}
~~~~~~~~~~\includegraphics[height=2.5in,width=2.5in]{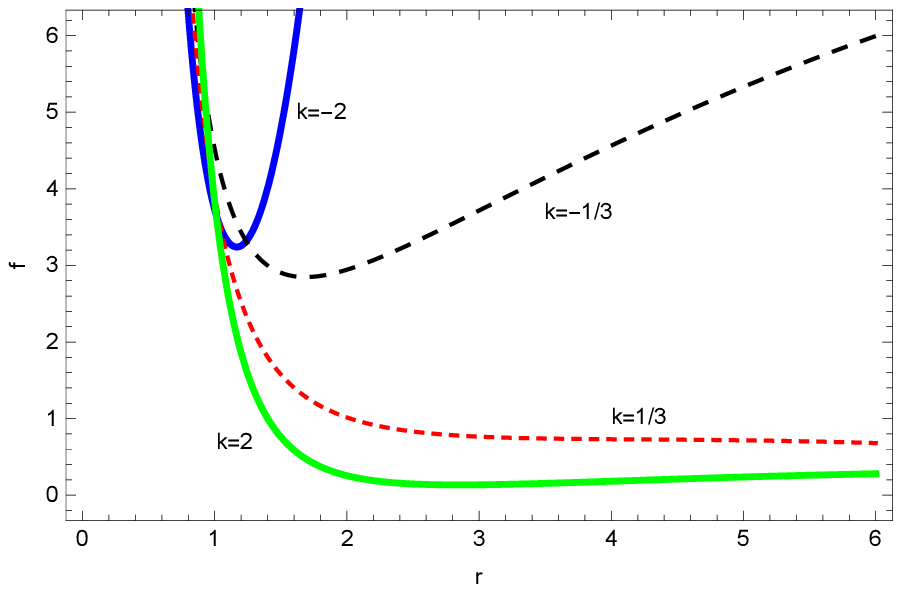}~~~~~~~~~~~~~\includegraphics[height=2.5in,width=2.5in]{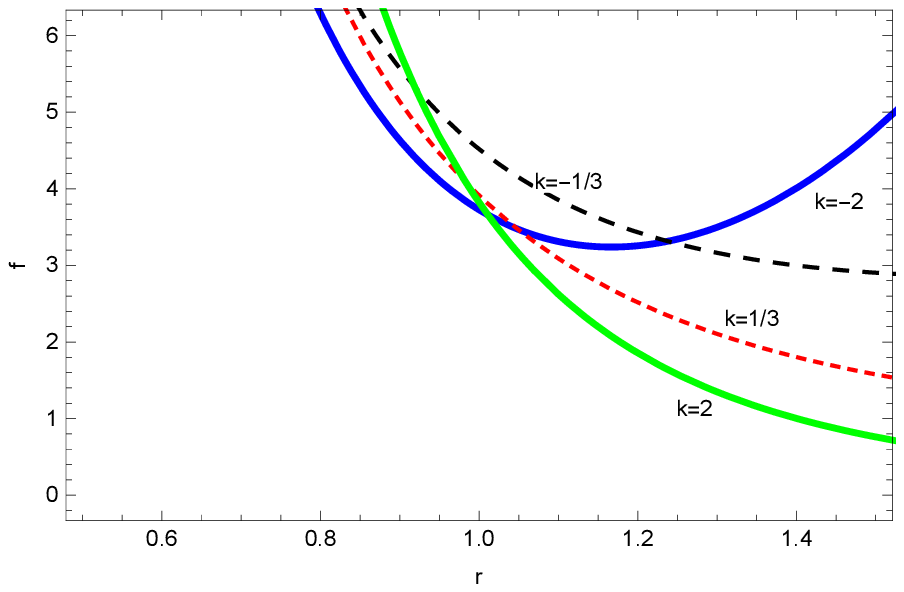}~~~~~~~\\\\

~~~~~~~~~~~~~~~~~~~~~~~~~~~~~~~Fig.6(a)~~~~~~~~~~~~~~~~~~~~~~~~~~~~~~~~~~~~~~~~~~~~~~~~~~~~~~~~~~~~Fig.6(b)~~~~~~~~\\

\vspace{2mm} \textit{\textbf{Figs 6 }show the variation of $f(t,r)$ against $r$
 for different values of $k$ in Brans-Dicke gravity's rainbow.\\\\
In fig.6(a) the other parameters are fixed at $\alpha=0.1, \beta=0.2$, $t=2$, $n=-1$, $\omega=-0.5$
$\gamma=5$, $V_{0}=0.1$, $\eta=1$, $E_{1}=1.42 \times 10^{-13}$,
$E_{p}=1.221 \times 10^{19}$. Fig. 6(b) represents the zoomed range of the outer horizon obtained in 6(a).}
\end{figure}

\begin{figure}
~~~~~~~~~~\includegraphics[height=2.5in,width=2.5in]{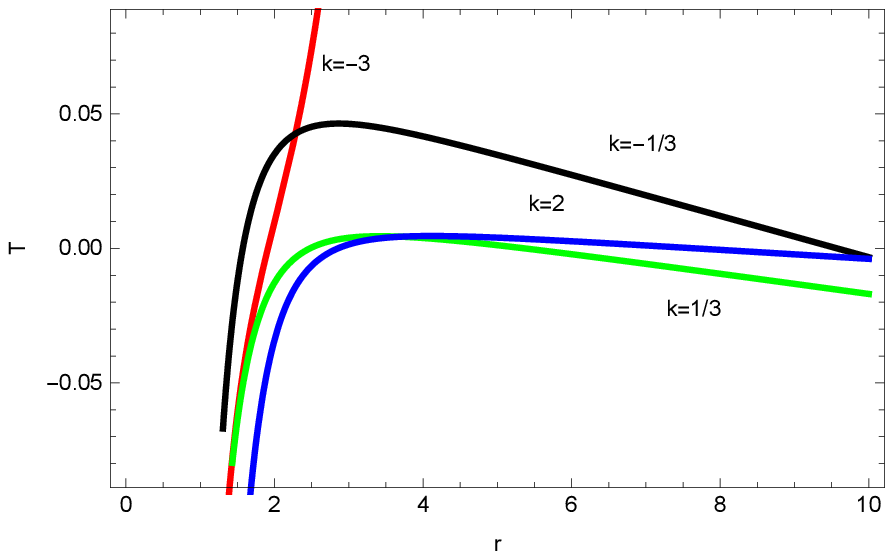}~~~~~~~~~~~~~\includegraphics[height=2.5in,width=2.5in]{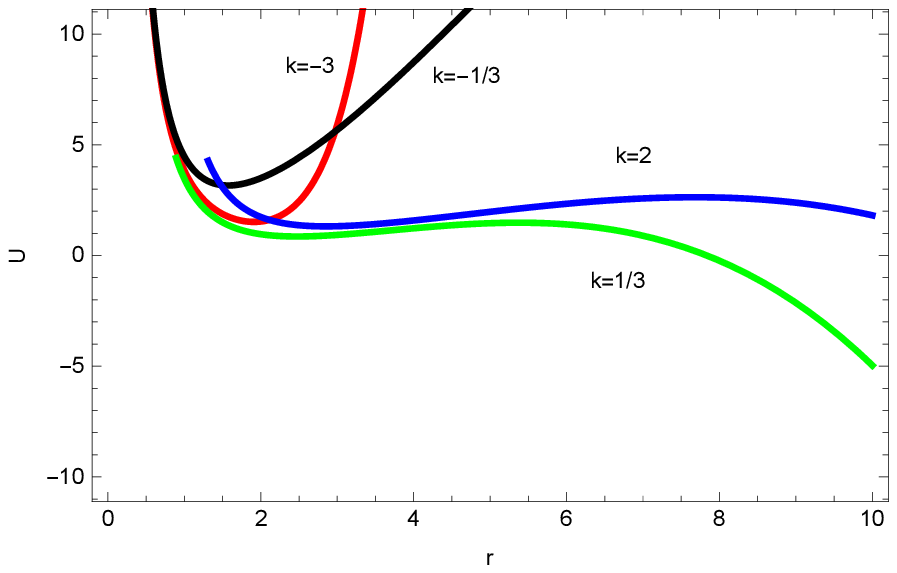}~~~~~~~\\\\

~~~~~~~~~~~~~~~~~~~~~~~~~~~~~~~Fig.7(a)~~~~~~~~~~~~~~~~~~~~~~~~~~~~~~~~~~~~~~~~~~~~~~~~~~~~~~~~~~~~Fig.7(b)~~~~~~~~\\

~~~~~~~~~~\includegraphics[height=2.5in,width=2.5in]{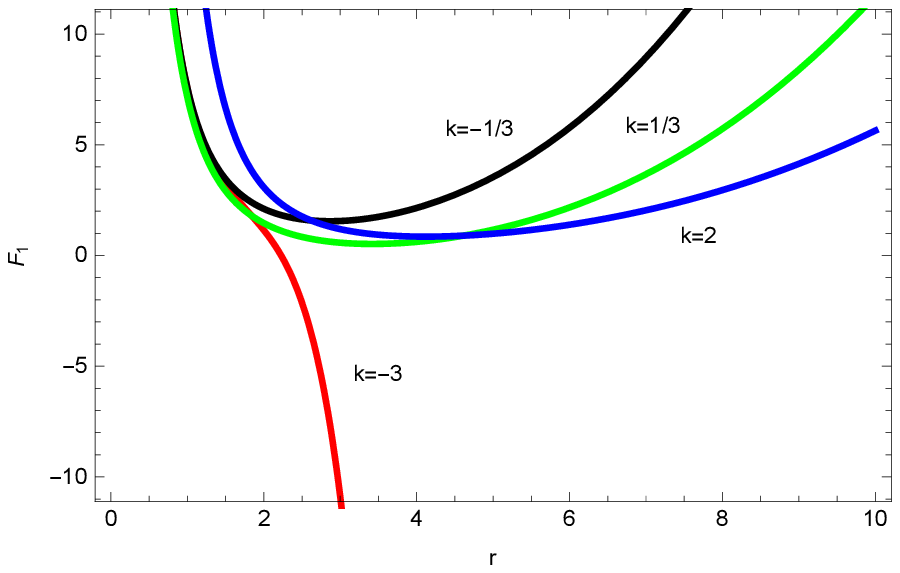}~~~~~~~~~~~~~\includegraphics[height=2.5in,width=2.5in]{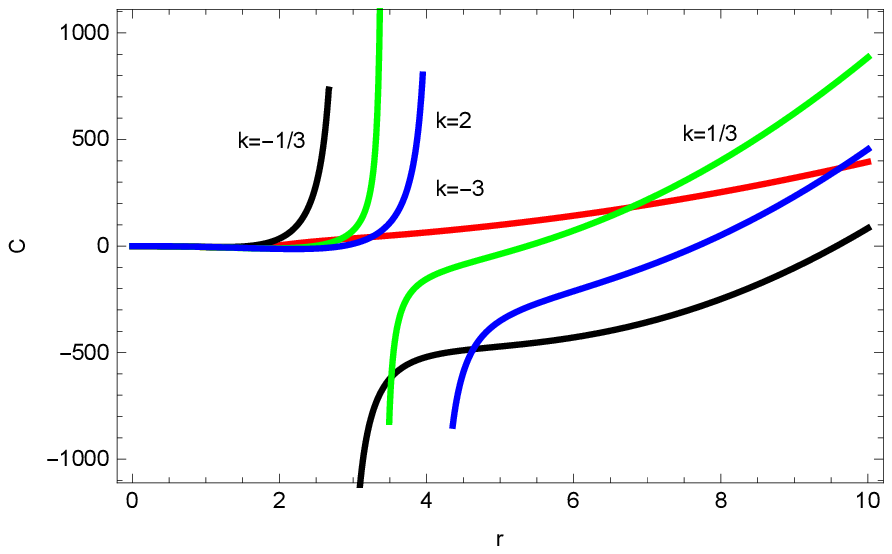}~~~~~~~\\\\

~~~~~~~~~~~~~~~~~~~~~~~~~~~~~~~Fig.7(c)~~~~~~~~~~~~~~~~~~~~~~~~~~~~~~~~~~~~~~~~~~~~~~~~~~~~~~~~~~~~Fig.7(d)~~~~~~~~\\

\vspace{2mm} \textit{\textbf{Figs 7} Figures 7(a), 7(b), 7(c), 7(d) show the variation of $T$, $U$, $F_{1}$ and $C$  against $r$ respectively
 for different values of $k$ in Brans-Dicke gravity's rainbow.\\\\
The other parameters are fixed at $\alpha=0.1, \beta=0.2$, $t=2$, $n=-1$, $\omega=-0.5$
$\gamma=5$, $V_{0}=0.1$, $\eta=1$, $E_{1}=1.42 \times 10^{-13}$,
$E_{p}=1.221 \times 10^{19}$. }
\end{figure}
\begin{figure}
~~~~~~~~~~\includegraphics[height=2.5in,width=2.5in]{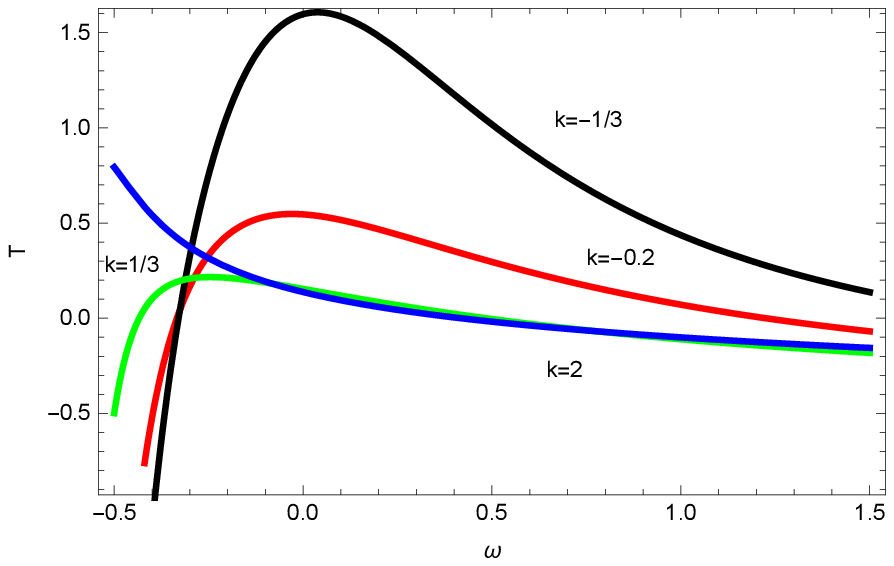}~~~~~~~~~~~~~\includegraphics[height=2.5in,width=2.5in]{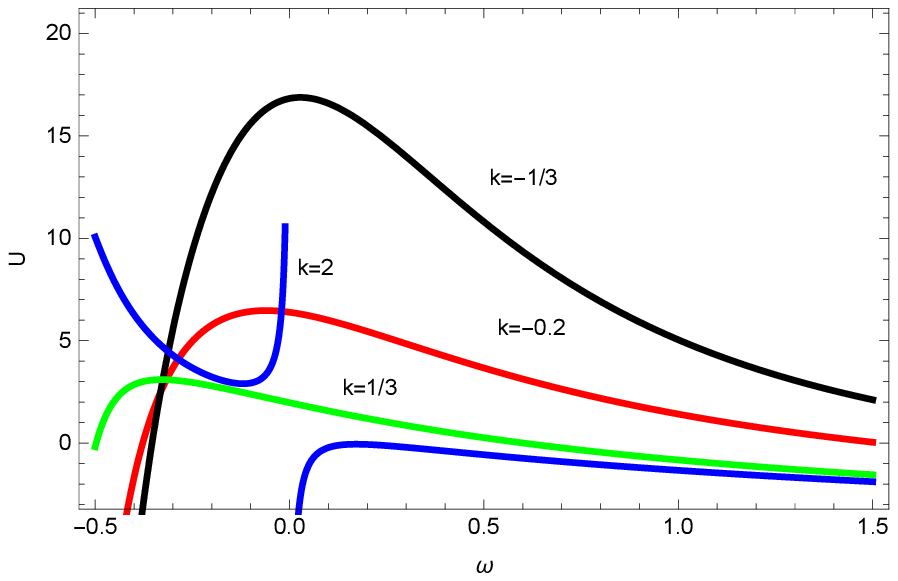}~~~~~~~\\\\

~~~~~~~~~~~~~~~~~~~~~~~~~~~~~~~Fig.8(a)~~~~~~~~~~~~~~~~~~~~~~~~~~~~~~~~~~~~~~~~~~~~~~~~~~~~~~~~~~~~Fig.8(b)~~~~~~~~\\

~~~~~~~~~~\includegraphics[height=2.5in,width=2.5in]{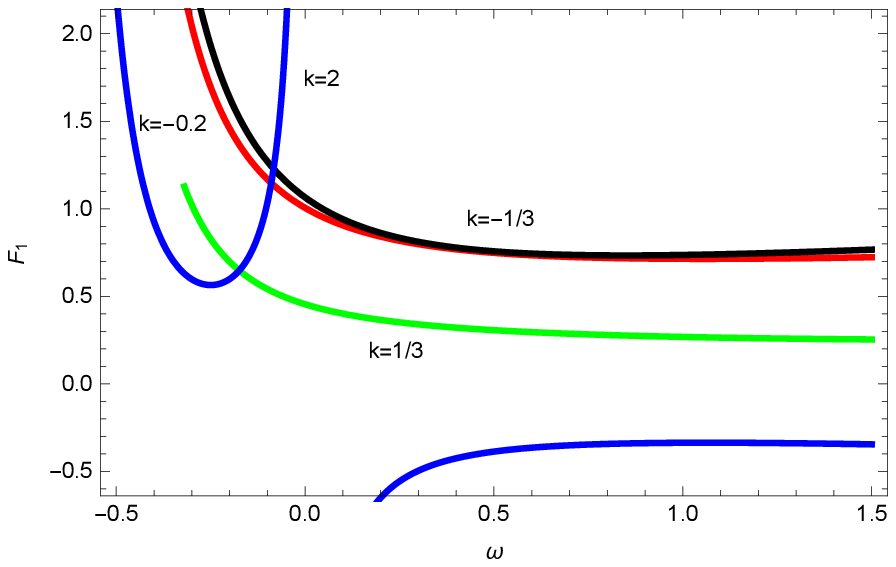}~~~~~~~~~~~~~\includegraphics[height=2.5in,width=2.5in]{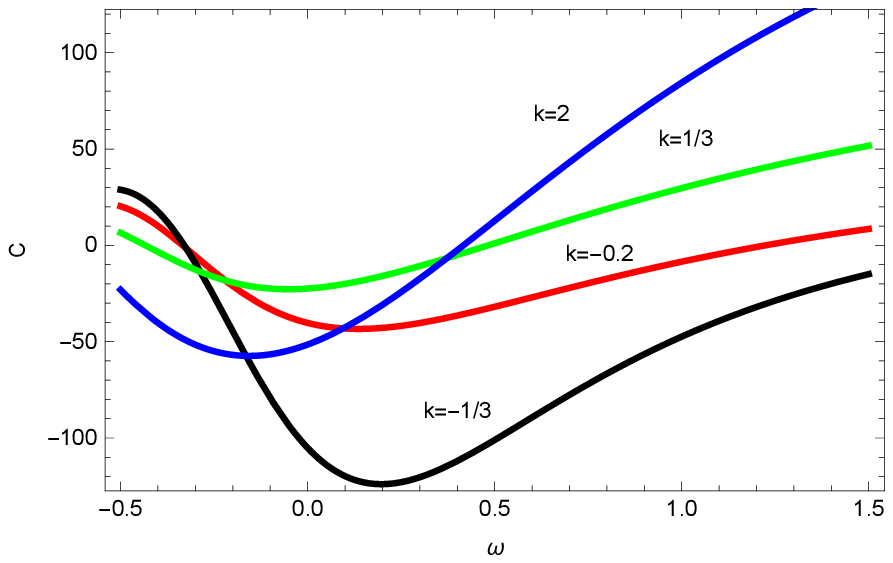}~~~~~~~\\\\

~~~~~~~~~~~~~~~~~~~~~~~~~~~~~~~Fig.8(c)~~~~~~~~~~~~~~~~~~~~~~~~~~~~~~~~~~~~~~~~~~~~~~~~~~~~~~~~~~~~Fig.8(d)~~~~~~~~\\

\vspace{2mm} \textit{\textbf{Figs 8} Figures 8(a), 8(b), 8(c), 8(d) show the variation of $T$, $F_{1}$, $U$ and $C$  against $\omega$ respectively
 for different values of $k$ in Brans-Dicke gravity's rainbow.\\\\
The other parameters are fixed at $r_{h}\approx1$, $\alpha=0.1, \beta=0.2$, $t=10^{-1}$, $n=-1$,
$\gamma=5$, $V_{0}=0.1$, $\eta=1$, $E_{1}=1.42 \times 10^{-13}$,
$E_{p}=1.221 \times 10^{19}$. }
\end{figure}
Setting the BD parameter $\omega=-0.5$ and all the other parameters as chosen earlier, we have plotted the termalization temperature $T$ against $r$ for four different value of $k$. It is observed that for $k=-1/3$, $k=1/3$ and $k=2$ at first the temperature is increasing function of $r$. The maximum temperature will occur in the region $(r_{h}-0.5,r_{h}+0.5)$ ($r_{h}\approx1$) and the temperature is decreasing for the radius $r>r_{h}$. For $k=-3$, the temperature is increasing function of $r$. In figure 8(a), $T$ is plotted against the BD parameter $\omega$ for $r_{h}\approx1$. It is observed that for the case of $\omega<0$, $T$ increases as $\omega$ increases and for $\omega>0$, $T$ decreases as $\omega$ increases.
The entropy is given by
\begin{equation}\label{S}
S=\pi^{2} r_{h}^{2},
\end{equation}
where we take $\pi G=1$. Consequently the total energy can be obtained from the relation
\begin{equation}\label{U}
U=\int T dS.
\end{equation}
Using equations (\ref{temperature}) and (\ref{S}), equation
(\ref{U}) yields the expression of total energy as follows
\begin{eqnarray*}\label{Unew}
U=\frac{\pi}{12\mathcal{F}(E)}\Big{[}\frac{6r^{\omega_{1}}(\omega_{1}-1)}{\omega_{1}}-\frac{r^{3}(1+\omega)^{-\frac{1}{1+\omega}}}{\mathcal{G}(E)f_{2}(t)^{2}
(\omega_{1}-3)(\omega_{2}-3)}\Big{(}-\frac{6kV_{0}f_{2}(t)^{n+1}(2+n+2\omega)r^{n}(1+\omega)^{\frac{n}{1+\omega}}}{(3+n+3\omega)}
\end{eqnarray*}
\begin{equation}
-\frac{3(\omega_{1}-3)(\omega_{2}-3)(-2k\omega+(1+\omega)^{\frac{1}{1+\omega}}(1+k-(1-k)\omega
f_{2}(t)))}{r(1+\omega)(\omega_{1}-2)(\omega_{2}-2)}
\dot{f_{2}(t)}+8(1+\omega)^{-\frac{1}{1+\omega}}\Big{)}+\frac{6f_{4}(t)(\omega_{2}-1)r^{\omega_{2}}}{\omega_{2}}\Big{]}.
\end{equation}
Figures 7(b) and 8(b) display the typical behavior of $U$ against $r$ and $\omega$ respectively for four different values of $k$.
Here we find that for $\omega>0$, the internal energy decreases as $\omega$ increases.\\

Another important thermodynamical quantity is the Helmholtz free energy, which reads
\begin{equation}\label{F}
F_{1}=U-TS.
\end{equation}
Exploiting equations (\ref{temperature}), (\ref{S}) and
(\ref{Unew}), equation (\ref{F}) yields
\begin{eqnarray*}
F_{1}=\frac{\pi}{12\mathcal{F}(E)}\Big{[}-\frac{3}{\omega_{2}}(\omega_{2}-1)(\omega_{2}-2)r^{\omega_{2}}f_{4}(t)-\frac{1}
{\mathcal{G}(E)(\omega_{1}-3)(\omega_{2}-3)}\Big{(}-\frac{4}{f_{2}(t)^{2}}r^{3}(1+\omega)^\frac{-2}{1+\omega}
\end{eqnarray*}
\begin{equation}
+3kV_{0}f_{2}(t)^{n-1}r^{n+3}(1+\omega)^{-1+\frac{(n-1)}{1+\omega}}\frac{(1+n+\omega)(2+n+2\omega)}{(3+n+3\omega)}\Big{)}-3r^{\omega_{1}}f_{3}(t)
(\omega_{1}-2)(1-\frac{1}{\omega_{1}})\Big{]}.
\end{equation}
Figure 7(c) and 8(c) represent the typical behavior of $F_{1}$ in terms of $r$ and BD parameter $\omega$ respectively considering different era of the evolution of the universe.
Finally we have considered the specific heat in constant volume
\begin{equation}
C=\left(\frac{dU}{dT}\right)_{V}.
\end{equation}
The above expression yields
\begin{eqnarray*}
C=2\pi^{2}(1+\omega)^{2}f_{2}(t)^{2}(\omega_{1}-3)(\omega_{2}-3)\mathcal{G}(E)\Big{[}r^{\omega_{1}-1}f_{3}(t)(\omega_{1}-1)+r^{\omega_{2}-1}f_{4}(t)(\omega_{2}-1)
\end{eqnarray*}
\begin{eqnarray*}
-\frac{r}{\mathcal{G}(E)f_{2}(t)^{2}}\Big{(}2(1+\omega)^\frac{-2}{1+\omega}
\Big{(}\frac{3r}{(\omega_{1}-3)(\omega_{2}-3)}+\frac{k\omega\dot{f_{2}}(t)(1+\omega)^{-1+\frac{1}{1+\omega}}}{(\omega_{1}-2)(\omega_{2}-2)}\Big{)}
-f_{2}(t)\Big{(}\frac{(1+k+(k-1)\omega)\dot{f_{2}}(t)}{(1+\omega)(\omega_{1}-2)(\omega_{2}-2)}
\end{eqnarray*}
\begin{eqnarray*}
+\frac{k
V_{0}r^{n+1}(2(1+\omega)+n)(1+\omega)^{-1+\frac{n-1}{1+\omega}}}{(\omega_{1}-3)(\omega_{2}-3)}\Big{)}\Big{)}\Big{]}\Big{[}k
V_{0}(1+\omega+n)(2(1+\omega)+n)f_{2}(t)^{n+1}r^{n}(1+\omega)^{\frac{n-1}{1+\omega}}
\end{eqnarray*}
\begin{equation}
-4(1+\omega)^{\frac{2\omega}{1+\omega}}+\mathcal{G}(E)f_{2}(t)^{2}(\omega_{1}-3)(\omega_{2}-3)(1+\omega)^{2}\left(f_{3}(t)(\omega_{1}-2)(\omega_{1}-1)
r^{\omega_{1}-3}+f_{4}(t)(\omega_{2}-2)(\omega_{2}-1)r^{\omega_{2}-3}\right)\Big{]}^{-1}
\end{equation}
In figure 7(d) and 8(d), we can observe the variation of specific
heat against the radius $r$ and BD parameter $\omega$
respectively. It is observed that the specific heat is taking
positive and negative values like any other thermodynamical
system. These thermal fluctuations lead to some instability in the
system with possible phase transition. Such instabilities get
corrected due to the presence of thermal fluctuations. There has
been a lot of studies done in this direction
\cite{1BPMF16,1JSBPMR16,2BPMF16,1BPMFUD16,1MFBP15,1BPMF15}. Also
from figure 8(d) we can conclude that specific heat is an
increasing function of the BD parameter $\omega$ in the region
$\omega>0$ for almost all chosen values of $k$ .

\section{Conclusions and Discussions}

In this note we have studied an energy dependent modification of a
time dependent geometry in the background of Brans-Dicke gravity
theory. The time dependent Vaidya metric representing a realistic
star was modified by rainbow functions in Brans-Dicke gravity. The
necessary field equations were formed and a solution was found. We
studied a gravitational collapse phenomenon under such conditions
to characterize the system. The concept of the existence of
outgoing radial null geodesics was used to explore the nature of
the gravitational singularity formed due to the collapse. The
existence of such outgoing geodesics from the central singularity
confirms the singularity to be a naked one. The absence of such
geodesics would indicate that the singularity is a black hole. In
our analysis we have considered the effects of both the graviton
mass as well as the rainbow deformations for the given time
dependent system. We have performed numerical simulations and
checked the nature of singularity by setting different initial
conditions. In all such cases we performed our analysis in the
late universe $(k<-1/3)$, i.e. a universe driven by dark energy.
In our study we have seen that under various scenarios the
singularity formed is a naked one. This is a significant
counter-example of the cosmic censorship hypothesis. We have also
checked the strength of singularity and obtained the conditions
under which the singularity can be called a strong singularity.

Lastly we have studied the thermodynamical behavior of this system
considering some important thermodynamical quantities. It is
observed that BD parameter $\omega$ affect those thermodynamic
quantities. For the case of $\omega<0$, thermalization temperature
$T$ increases as $\omega$ increases and for $\omega>0$, $T$
decreases as $\omega$ increases. The internal energy and specific
heat have also been studied and it is found that for $\omega>0$,
the internal energy decreases as $\omega$ increases. For some
special values of $k$ we have seen some instability with possible
phase transition.

\section*{Acknowledgments}

PR acknowledges University Grants Commission, Govt. of India for
providing research project grant (No.F.PSW-061/15-16 (ERO)). PR
also acknowledges the Inter University centre for astronomy and
astrophysics(IUCAA), Pune, India for granting visiting
associateship.


\end{document}